\documentclass[aps,prb,twocolumn, showpacs,superscriptaddress,floatfix]{revtex4}
\usepackage{graphicx}
\usepackage{amsmath}
\usepackage{amsthm}
\usepackage{multirow}
\usepackage{relsize}

\usepackage{epstopdf}
\newcommand{\BEQ}{\begin{equation}}
\newcommand{\EEQ}{\end{equation}}
\newcommand{\BEA}{\begin{eqnarray}}
\newcommand{\EEA}{\end{eqnarray}}

\begin{document}

\title{Infinite volume extrapolation in the one-dimensional bond diluted Levy
  spin-glass model near its lower critical dimension}

\author{L.~Leuzzi}
\affiliation{IPCF-CNR, UOS Roma Kerberos, Universit\`a La Sapienza, P. le A. Moro 2, I-00185, Rome, Italy} 
\author{G.~Parisi}
\author{F.~Ricci-Tersenghi}
\affiliation{Dipartimento di Fisica and INFN--Sezione di Roma1, Universit\`a La Sapienza, P. le A. Moro 2, I-00185, Rome, Italy}
\affiliation{IPCF-CNR, UOS Roma Kerberos, Universit\`a La Sapienza, P. le A. Moro 2, I-00185, Rome, Italy} 
\author{J.~J.~Ruiz-Lorenzo} \affiliation{Departamento de
  F\'{\i}sica and 
Instituto de Computaci\'on Cient\'{\i}fica Avanzada (ICCAEx),
  Universidad de Extremadura, 06071 Badajoz, Spain.}
\affiliation{Instituto de Biocomputaci\'on y
  F\'{\i}sica de Sistemas Complejos (BIFI), 50018 Zaragoza, Spain.}

\date{\today}

\begin{abstract}

We revisited, by means of numerical simulations, the one dimensional
bond diluted Levy Ising spin glasses outside the limit of validity of
mean field theories.  In these models the probability that two spins
at distance $r$ interact (via a disordered interactions, $J_{ij}=\pm
1$) decays as $r^{-\rho}$.  We have estimated, using finite size
scaling techniques, the infinite volume correlation length and spin
glass susceptibility for $\rho=5/3$ and $\rho=9/5$.  We have obtained
strong evidence for divergences of the previous observables at a non
zero critical temperature.  We discuss the behavior of the critical
exponents, especially when approaching the value $\rho=2$,
corresponding to a critical threshold beyond which the model has no
phase transition. Finally, we numerically study the model right at 
the threshold value $\rho=2$.
\end{abstract}


\maketitle

\section{Introduction and model definition}

The study of spin glasses on finite dimensional lattices is a notoriously
difficult problem, because of very strong finite size effects.
Recently, there has been a renewed interested in long range models, since these allow
to interpolate between mean-field critical behavior and finite dimensional
one.  In particular, the diluted version of these long range models is very
efficient and allows us to simulate very large sizes, thus reducing the finite
size effects.

In the present work we show that for these diluted models with long range
interactions it is actually possible to extract the asymptotic scaling
functions from the numerical data. These functions allow us to estimate the
values of the observables in the thermodynamic limit, and consequently to
estimate the critical behavior in an alternative way than the one already used
in Ref.~\onlinecite{Leuzzi08}.  In addition, we want to extract the elusive
correction-to-scaling exponent to compare with previous numerical
computations.~\cite{Leuzzi08}

Another interesting issue is the understanding of the behavior of the Ising
spin glass model near its lower critical dimension focusing on the breakdown
of scaling laws (e.g. via logarithmic corrections). In the Edward-Anderson
model, the lower critical dimension was estimated in Ref. \onlinecite{Franz}
as $D_{\mathrm L}=2.5$ and recently it has been studied experimentally in
(thin) spin glass films.~\cite{Orbach} We can also study this issue in the
long range model by tuning the power law decay exponent of the couplings.

We study the one dimensional Ising spin glass ($\sigma_i=\pm 1$) with
Hamiltonian \cite{Leuzzi08,Leuzzi09}
\begin{equation}
{\cal H}=-\sum_{i<j} J_{ij} \sigma_i\sigma_j\,\,.
\label{eq:ham}
\end{equation}
The quenched random couplings $J_{ij}$ are independent and identically
distributed random variables taking a non zero value with a
probability decaying with the distance between spins $\sigma_i$ and
$\sigma_j$, $r_{ij}\equiv \min(|i-j|,L-|i-j|)$, as
\begin{equation}
\mathbf{P}[J_{ij}\neq 0] \propto r_{ij}^{-\rho}\quad \text{for}\;\; r_{ij}
\gg 1\;.
\label{eq:Jij}
\end{equation}
Non-zero couplings take value $\pm 1$ with equal probability.  We use
periodic boundary conditions and a $z=6$ average coordination number.  

We will briefly review the most important characteristic of this
model. The most important point is that the $\rho$ parameter
determines the universality class of the model.  In Table
\ref{tab:rho} the different critical behaviors as a function of the
value of $\rho$ are reported.

\begin{table}[b!]
\centering
\begin{tabular}{||c|c|c||}
\hline
$\rho$ & $D(\rho)$ &transition type\\
\hline
$\leq 1$ &   $\infty$  &Bethe lattice like\\
$(1,4/3]$ & $[6,\infty)$ & $2^{\rm nd}$ order, MF\\
$(4/3,2]$ & $[2.5,6)$ & $2^{\rm nd}$ order, non-MF\\
$2$ & $2.5$ &Kosterlitz-Thouless or $T=0$-like\\
$>2$ & $<2.5$ &none \\
\hline 
\end{tabular}
\vspace{.2 cm}
\protect\caption{ From infinite range to short range behavior of the
  SG model defined in Eqs.(\ref{eq:ham},\ref{eq:Jij}).}
\label{tab:rho}
\end{table}

For $\rho > 1$ the critical behavior turns out to be equal to the one of the
fully connected version of the model,~\cite{FT_LR} where bonds are Gaussian
distributed with zero mean and a variance depending on the distance as
${\overline{J_{ij}^2}} \propto r_{ij}^{-\rho}$. By changing $\rho$, the model
displays different behaviors:~\cite{Leuzzi08} for $\rho
\le\rho_\mathrm{U}\equiv 4/3$, the mean-field (MF) approximation is exact,
while for $\rho > \rho_\mathrm{U}$, infrared divergences arise and the MF
approximation breaks down.  The value $\rho_{\rm U}=4/3$ marks the equivalent
of the upper critical dimension of short-range spin-glasses in absence of an
external magnetic field ($D_{\rm U}=6$).  At $\rho>\rho_{\rm L}=2$ no finite
temperature transition occurs, even for zero magnetic field, $h=0$.~\cite{Campanino87} A
relationship between $\rho$ and the dimension $D$ of short-range models can be
expressed as $\rho=1+2/D$ which is exact at $D_{\rm U}=6$ ($\rho_{\rm U}=4/3$)
and approximated as $D<D_{\rm U}$. Indeed, according to this analogy, the
lower critical dimension $D_{\rm L} \simeq 2.5$(see Refs. \onlinecite{Franz}
and \onlinecite{Boettcher05}) would correspond to $\rho \simeq 1.8$, rather
than to $\rho=\rho_{\rm L}=2$.  An improved equation relating short-range
dimensionality $D$ and power-law long-range exponent $\rho$ includes the value
of the critical exponent of the space correlation function for the short-range
model, $\eta(D)$, and reads:~\cite{KYultimo}
\begin{equation} 
\rho(D)=1+\frac{2-\eta(D)}{D}\,. 
\end{equation}
In systems whose lower critical dimension is not fractional and
$\eta(D_{\rm L})$ can be explicitely estimated the above relationship
guarantees, at least, that $\rho(D_{\rm L}) = \rho_{\rm L}$, though
some discrepancies have been observed, as well, in between $\rho_{\rm
  U}$ and $\rho_{\rm L}$, see, e.g.,
Refs. [\onlinecite{Banos12,Ibanez13,Leuzzi13, Angelini14}].

A large number of studies concentrated on the parameter region around
the threshold between mean-field-like behavior and non-mean-field one
($\rho_\mathrm{MF}=4/3$).  The present work focuses, instead, on large values
of $\rho$ ($\rho=5/3,9/5, 2$), whose critical behavior is similar
to the behavior of short-range interacting models in low dimension,
close to the lower critical one.  As expected in general for low
dimensional systems, these models show more severe finite size effects
than previously studied cases.  The aim of the present analysis is to
show that a faithful extrapolation of the critical behavior in the
thermodynamic limit can be achieved also in these harder cases, by
means of improved finite size scaling techniques.  These techniques
are based on those developed in Ref.
[\onlinecite{Caracciolo95}] and involve the estimate of the leading
correction-to-scaling exponent. In this paper we will provide a
comprehensive study of these scaling correction tackling with the
confluent (analytical) corrections and the non-confluents ones.

Finally, a further motivation for this numerical study is the comparison with
an analytical estimate of the divergence of the correlation length in the
$\rho=2$ model obtained by Moore,\cite{Moore10} (for $\rho=2$ the model is at
its lower critical dimension). In addition, we are interested to research
possible logarithmic corrections to the scaling laws just at the lower
critical dimension.

\section{Observables and the Finite Size Scaling Method}

The onset of spin glass long range order can be studied using 
the four-point correlation function
\begin{equation}
C(x)=\sum_{i=1}^L{\overline{\langle \sigma_i\sigma_{i+x}\rangle^2 }}
\end{equation}
where indices should be intended modulo $L$ and we have denoted the average over quenched disorder by
$\overline{(\cdot\cdot\cdot)}$ and the thermal average by $\langle
(\cdot\cdot\cdot) \rangle$.
In terms of Fourier transform $\tilde C(k)$ one can express
both the SG susceptibility
\begin{equation}
\chi_{\rm sg} \equiv \tilde C(0) 
\label{eq:TD_chi}
\end{equation}
and the so-called second-moment
correlation length\cite{Caracciolo95}
\begin{equation}
\xi_2 \equiv \frac{L}{2 \pi}\left[\frac{\tilde C(0)}{\tilde
    C(2\pi/L)}-1\right]^{\frac{1}{\rho-1}}\;
\label{eq:xiL}
\end{equation}
Notice that, for
the simulated lattice sizes $\sin(\pi/L)\simeq \pi/L$.

We will describe in the next paragraphs the Finite Size Scaling (FSS)
method that we have used to analyze the data.~\cite{Caracciolo95} Consider
a singular observable $O$ diverging at the critical temperature $T_c$
as $|T-T_c|^{-y_O}$. Discarding corrections to scaling, we can
write
\begin{equation}
\frac{O(T,L)}{O(T,\infty)}=f_O\left(\frac{\xi_2(T,\infty)}{L}\right)\,,
\label{FSS1}
\end{equation}
being $f_O(x)$ an universal function, decaying at large $x$ as $f_O(x) \sim
x^{-y_O/\nu}$.  For the observables of our interest, i.e., spin glass
susceptibility and correlation length, we have $y_\chi =
\gamma$ and $y_{\xi_2} = \nu$ and, therefore,
\begin{eqnarray}
f_\chi(x)  &\sim& x^{-\gamma/\nu} = x^{1-\rho}
\label{FSS2} \\
f_{\xi_2}(x) &\sim& 1/x \qquad \qquad\qquad \text{for } x \to \infty
\nonumber
\end{eqnarray}
where we have used the fact the $\eta$ exponent does not
renormalize in long-range systems and takes the value $\eta=3-\rho$.

From Eq. (\ref{FSS1}), we can write, as well,
\begin{equation}
\frac{O(T,2 L)}{O(T,L)}=F_O\left(\frac{\xi_2(T,L)}{L}\right)\,,
\label{FSS3}
\end{equation}
where $F_O$ is another universal function.

To extrapolate our measures to infinite volume, we have followed the
procedure described in Refs. \onlinecite{Caracciolo95} and
\onlinecite{Palassini99}.  We perform Monte Carlo simulations on
different pairs $(T,L)$ computing generic observables, $O(T,L)$, among
which, in particular, the correlation length, $\xi_2(L,T)$.  This
allows us to plot $O(T, 2 L)/O(T,L)$ against $\xi_2(T,L)/L$: if all
the points lie on the same curve, Eq.~(\ref{FSS3}) holds and the
scaling corrections are negligible.  We can, thus, compute the scaling
functions $F_O$ and $F_{\xi_2}$. From these we can iteratively
extrapolate the infinite volume pair $(\xi_2,O)$.  In our simulations
we approach the $L\to\infty$ limit along the sequence $L \to 2 L \to 2^2
L \to \cdot \cdot \cdot \to \infty$.  In order to do such an
extrapolation we need a smooth interpolating function for $F_O(z)$.

For short range models, previous studies~\cite{Caracciolo95,Palassini99}
 used interpolating functions of the kind
\begin{equation}
F_O(x)=1+\sum_{k=1}^n a_k^O \exp(-k/x)\,,
\label{OLD}
\end{equation}
where the coefficients $a_k^O$ depend on the observable $O$ and,
typically, $n\simeq 4$. This functional form was based on the theory
of the two dimensional $O(3)$ model~\cite{Caracciolo95} and worked
satisfactorily in the three dimensional Ising spin
glass.~\cite{Palassini99}

In the present case Eq. (\ref{OLD}) does not interpolate
well the numerical data and we need to resort to a different functional form.
We have, thus, introduced the following parameterization of the scaling
functions $F_{\xi_2}$ and $F_\chi$:
\begin{equation}
F_O(z)= 1+\frac{a_1 z}{a_2+z}+\frac{a_3 z}{a_4+z}\,.
\label{scaling}
\end{equation}
where the $a$'s coefficients depend on the choice of the observable
$O$ and for $O=\xi_2$ they must satisfy the constraint $a_1+a_3=1$.
This parameterization works really well for all values of $\rho$ and
for both the measured correlation length and spin glass
susceptibility. In the $\rho=2.0$ case we have used,  as well, 7th and 8th
degree cubic spline polynomial fits\cite{NUMREC} to compare with the new
interpolation proposed.

\section{Numerical Simulations}

We have simulated the model using the Metropolis algorithm and
multi-spin coding (we have simulated 64 systems in parallel). In
addition, to thermalize samples in the low temperature region we have
used the parallel tempering method.~\cite{PT} In order to check
thermalization we have looked at the temporal evolution of each
observable measured on a logarithmic time scale.  In Table
\ref{tab:NS} we report all the parameters used in our
simulations. As a control, we have also simulated small lattices.

\begin{table}[b!]
\centering
\begin{tabular}{|c|c|c|c|}
\hline
 &\multicolumn{3}{ |c| }{$N_s$} \\
\hline
$B$ & $\rho=5/3$ & $\rho=9/5$ & $\rho=2$ \\ \hline
6   & 154624     & 119808     & 64000 \\
7   & 113536     &            & 320000 \\
8   & 39936      & 178688     & 6656 \\
9   & 21632      & 33792      & 45056 \\
10  & 29824      & 154368     & 12544 \\ 
11  & 54912      & 92160      & 24064\\  
12  & 38784      & 38912      &      \\
13  & 16512      &            &       \\
\hline \hline
$[T_m, T_M]$  & $[1.4,2.8]$  & $[1.1, 2.2]$ & $[0.5, 2.3]$ \\
$\Delta T$    & 0.05                & 0.05              & 0.05 \\
\hline
\end{tabular}
\protect\caption{Parameters of the numerical simulations. $B\equiv \log_2 L$,
  $N_s$ is the number of samples,  and $T_m$, $T_M$ and $\Delta T$ are the
  lowest temperature, the highest one and the temperature step in the parallel
  tempering method.}
\label{tab:NS}
\end{table}

\section{Numerical Results for the Critical Behavior}

\subsection{Critical behavior for $\rho=5/3$ ($3<D<4$)}

\begin{figure}[t!]
\centering
\includegraphics[width=0.66\columnwidth, angle=270]{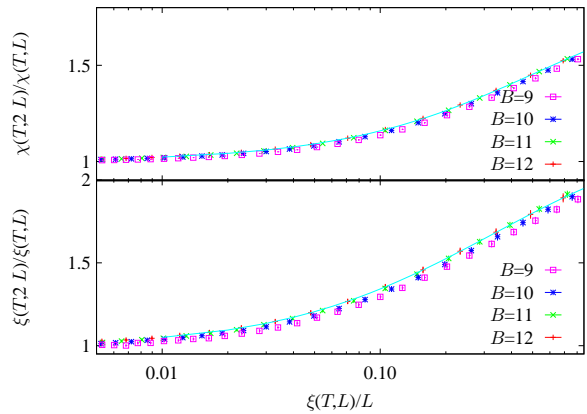}
\caption{(color online) Rescaled data for $\rho=5/3$ according to Eq.~(\ref{FSS3})
  and interpolating FSS functions $F_\chi$ (top) and $F_{\xi_2}$
  (bottom) for sizes $L=2^B$. Error bars are smaller than symbol
  sizes. The two continuous lines are the fits performed with
  Eq. (\ref{scaling}). In the figures of this paper we will use $\xi$
  as $\xi_2$.}
\label{fig:fig1r1_666}
\end{figure}

In Fig.~\ref{fig:fig1r1_666} we test the Finite Size Scaling Ansatz in the
form of Eq. (\ref{FSS3}). We can still see weak scaling corrections for the
smallest plotted value of the lattice size ($2^{9}$), but all data for larger
 sizes lie on the same curves both for the susceptibility
(top panel of Fig.~\ref{fig:fig1r1_666}) and the correlation length (bottom
panel of Fig.~\ref{fig:fig1r1_666}). The next step is to interpolate the data
with the scaling function defined in Eq.~(\ref{scaling}).  The fit is
good with a $\chi^2/\mathrm{d.o.f.}$ equal to $5.1/26$ and $3.2/25$ for
the $\xi_2$ and $\chi$, respectively (discarding the $x$-error bars).
Statistical errors on the extrapolated observables ($\xi(T)$ and $\chi(T)$)
 are estimated using the same Monte Carlo technique introduced in Ref.~[\onlinecite{Caracciolo95}].

\begin{figure}[t!]
\centering
\includegraphics[width=.34\textwidth, angle=270]{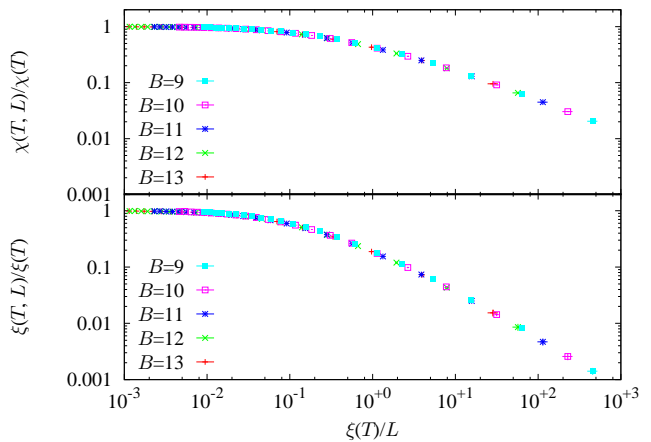}
\caption{(color online) Rescaled data for $\rho=5/3$ according to Eq.~(\ref{FSS1}).}
\label{fig:fig2r1_666}
\end{figure}

\begin{figure}[t!]
\centering
\includegraphics[width=.34\textwidth, angle=270]{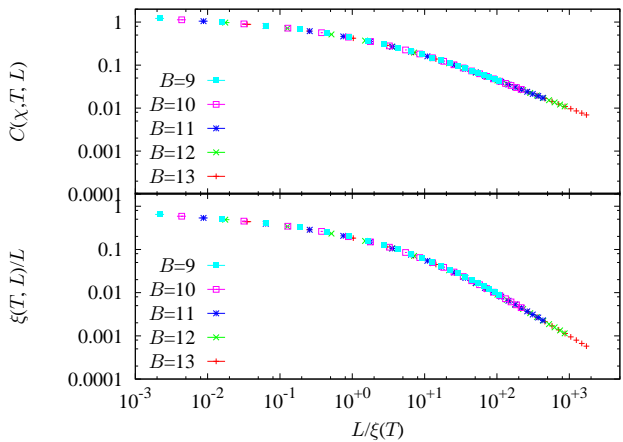}
\caption{(color online) Test on the scaling functions $f_\chi$ and $f_\xi$ for $\rho=5/3$. We
  plot $C(O,T,L) \equiv x^{y_O/\nu} f_O(x)$ versus $1/x\equiv L/\xi(T)$ ($O=\xi,
  \chi$). Notice that $C(\xi,T,L)=\xi(T,L)/L$. }
\label{fig:fig2r1_666b}
\end{figure}

Once we have the extrapolated values of $\xi_2$ and $\chi$, as a
consistency test, we check if Eq.~(\ref{FSS1}) holds. We present this
test in Fig.~\ref{fig:fig2r1_666}. We can see that all the points are
lying on the same universal curves corresponding to $f_\chi$ (top) and
$f_{\xi_2}$ (bottom).  For large $x$ a simple fitting procedure
returns $f_{\xi_2}(x)\sim x^{-0.91(5)}$ and $f_\chi(x) \sim
x^{-0.59(4)}$, not far from the behavior predicted Eq.~(\ref{FSS2}),
but nevertheless underestimating the exponent values.  However, data in
Fig.~\ref{fig:fig2r1_666} clearly show a downward bending, even for
the largest $\xi(T)/L$, thus suggesting that finite size effects still
prevent a proper asymptotic  estimate for the exponents (so, we need to take
into account scaling corrections).  
An improved test  can be obtained by plotting the quantity
\begin{equation}
C(O,L,T)\equiv x^{y_O/\nu} f_O(x)
\label{def:COLT}
\end{equation}
versus $1/x\equiv L/\xi(T)$, that is expected to extrapolate to a finite value
on the $y$ axis (as $L/\xi(T)\to 0$ or equivalently $\xi(T)\to \infty$). 
The results of this test clearly confirm this behavior, as
shown in Fig.~\ref{fig:fig2r1_666b}.

\begin{figure}[t!]
\centering
\includegraphics[width=.34\textwidth, angle=270]{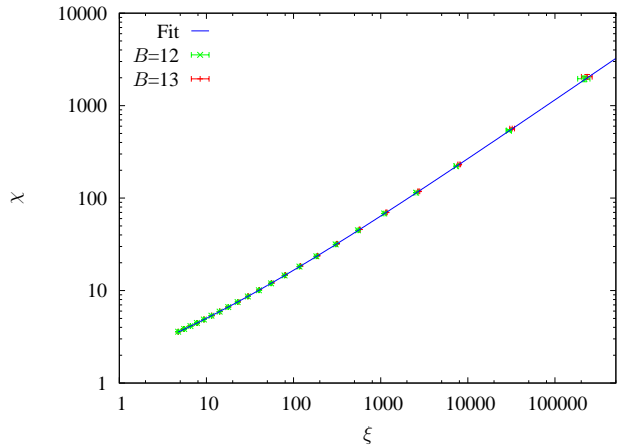}
\caption{(color online) Extrapolated thermodynamical susceptibility $\chi$ versus
  $\xi_2$ for $\rho=5/3$. The two data sets correspond to
  extrapolations obtained by using up to $B=13$ data (red points)
  and up to $B=12$ data (green points).}
\label{fig:fig3r1_666}
\end{figure}

Using the interpolating functions for $F_\chi$ and $F_{\xi_2}$ (see
Fig.~\ref{fig:fig1r1_666}) we can extrapolate susceptibility and correlation
length to the thermodynamic limit. In Fig.~\ref{fig:fig3r1_666} we show the
resulting infinite volume susceptibility, which is well fitted by the usual
power law, including scaling corrections,
\begin{equation}
\chi= A \xi_2^{2-\eta} \left(1+ B \xi_2^{-\Delta} \right)+C.
\label{fit_chi}
\end{equation}
Notice that the constant $C$ in the fit takes into account the background in the
susceptibility induced by the analytic part of the free energy.

Fitting in the range $\xi_2>10$ we obtain: $\eta=1.353(15)$ and
$\Delta=0.4(1)$ ($\chi^2/\mathrm{d.o.f.}=4.5/12$).  
\footnote{In order to understand the
  effect of discarding the error bars in $\xi_2$, we have performed a pure
  power law fit $ \chi= A \xi_2^{2-\eta} $ provides with $\eta=1.368(6)$ (with
  no $\xi_2$-errors) and by using the routine of Numerical
  Recipes\cite{NUMREC} which takes into account errors in $\xi_2$ as well as
  in $\chi$, $\eta=1.366(15)$ (in both cases the quality of the fit is really
  good). Hence, the error in $\eta$ has doubled.}  Eventually, we can exploit
the knowledge of the exact value $\eta=3-\rho=4/3$ and find a better estimate
for the correction-to-scaling exponent $\Delta=0.28(2)$
($\chi^2/\mathrm{d.o.f.}=5.4/13$).

\begin{figure}[t!]
\centering
\includegraphics[width=.34\textwidth, angle=270]{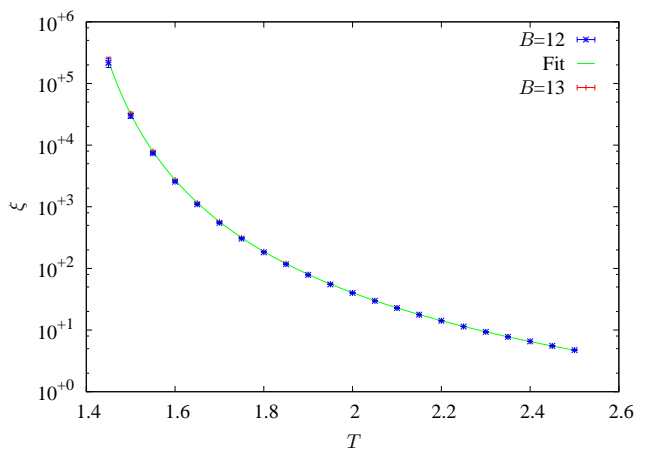}
\caption{(color online) $\rho=5/3$. Extrapolated thermodynamical correlation length $\xi_2$
  versus temperature, together with the best fit. The two data sets correspond
  to extrapolations obtained by using up to $B=13$ data (red points) and up to
  $B=12$ data (green points).}
\label{fig:fig4r1_666}
\end{figure}

The final step of the analysis is to compute the critical temperature $T_c$,
the correlation length exponent $\nu$, and the scaling correction exponent
$\theta$, according to the following equation
\begin{equation}
\xi_2(T,\infty) =A (T-T_c)^{-\nu} \left( 1+ B (T-T_c)^\theta  \right)\,.
\label{fit_xi}  
\end{equation}
By fitting the data in the range $T \le 2.3$ we
obtain $T_c=1.35(1)$, $\nu=5.0(3)$ and $\theta=1.9(1)$ with a
$\chi^2/\mathrm{d.o.f.}=5.4/13$, cf. Fig.~\ref{fig:fig4r1_666}.

If we associate $\Delta$ and $\theta$ to non-confluent scaling
corrections, one should have $\theta=\nu \Delta$. Taking the estimates
of $\nu$ and $\theta$ from the $\xi_2$-fit, we obtain
$\theta/\nu=0.38(3)$, which compares well with the values obtained for
$\Delta$ from the $\chi$ versus $\xi_2$ fit.

As an additional test of the extrapolation procedure, we show in
Figs.~\ref{fig:fig3r1_666} and \ref{fig:fig4r1_666} the infinite
volume results obtained using data from simulations of system sizes up
to $B=12$ (green points) and up to $B=13$ (red points), that coincide
very well within the errors.

Finally, we can compare the above results with previous estimates~\cite{Leuzzi08} obtained
using the quotient method:~\cite{quotient} $T_c=1.36(1)$, $\nu=5.3(8)$ and
$\omega=0.8(1)$.  While $T_c$ and $\nu$ agree well, the correction-to-scaling
exponent $\omega$ is different from the $\Delta$ exponent measured here. A
similar disagreement on the value of the correction to scaling exponent in
long range models has been recently observed in Ref.~\onlinecite{Angelini14}.

\subsection{Critical behavior for $\rho=9/5$ ($D_{\rm L}<D<3$)}

\begin{figure}[t!]
\centering
\includegraphics[width=.34\textwidth, angle=270]{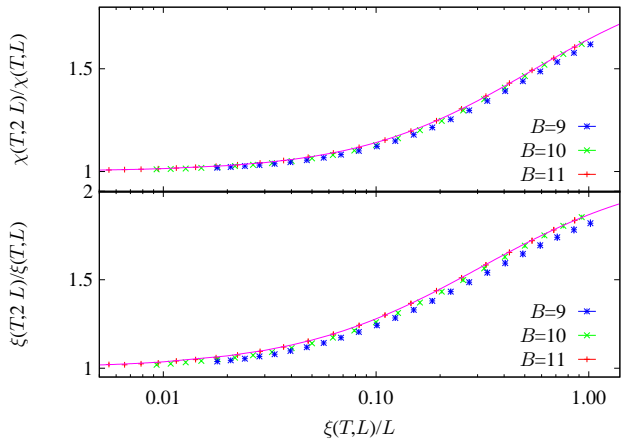}
\caption{(color online) FSS functions $F_{\xi_2}$ (below) and $F_\chi$ (above)for
  $\rho=9/5$. We plot data for lattice sizes $2^{B}$ with $B=9,10,11$
  and $12$. The error bars are less than the size of the symbols. The
  two continuous lines are the fits using Eq. (\ref{scaling}).}
\label{fig:fig1r1_8}
\end{figure}

We will be following the same procedure to extract the critical exponents
as described in the previous subsection.  In Fig. \ref{fig:fig1r1_8}
we test the Finite Size Scaling Ansatz in the form of
Eq. (\ref{FSS3}). Also for this value of $\rho$ all the data from different
lattice sizes, but the smallest one, lie on the same universal curve both for the
susceptibility (top panel) and the correlation length (bottom
panel). The next step is to parameterize the two universal functions
by means of a fit. The fits proposed in references
[\onlinecite{Caracciolo95},\onlinecite{Palassini99}] fail again for
this value of $\rho$. We have rather used that of Eq. (\ref{scaling})
for the interpolation, displaying a $\chi^2/\mathrm{d.o.f.}=15.7/17$ for the
susceptibility and $\chi^2/\mathrm{d.o.f.}=1.1/18$ for the correlation
length.

\begin{figure}[t!]
\centering
\includegraphics[width=.34\textwidth, angle=270]{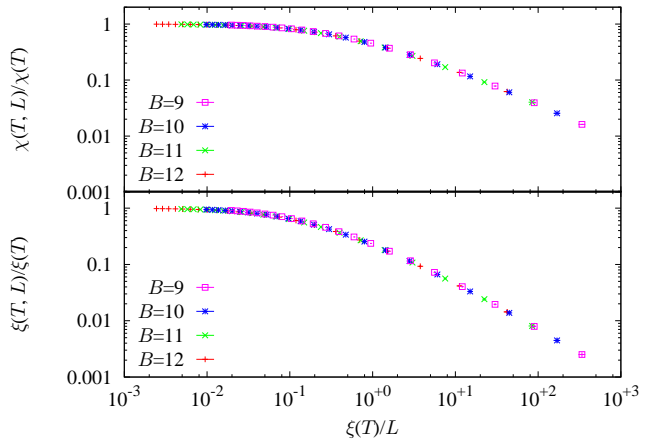}
\caption{(color online) FSS functions $f_{\xi_2}$ (below) and $f_\chi$ for $\rho=9/5$. 
  (above). The error bars are less than the size of the symbol}
\label{fig:fig2r1_8}
\end{figure}

\begin{figure}[h]
\centering 
\includegraphics[width=.34\textwidth, angle=270]{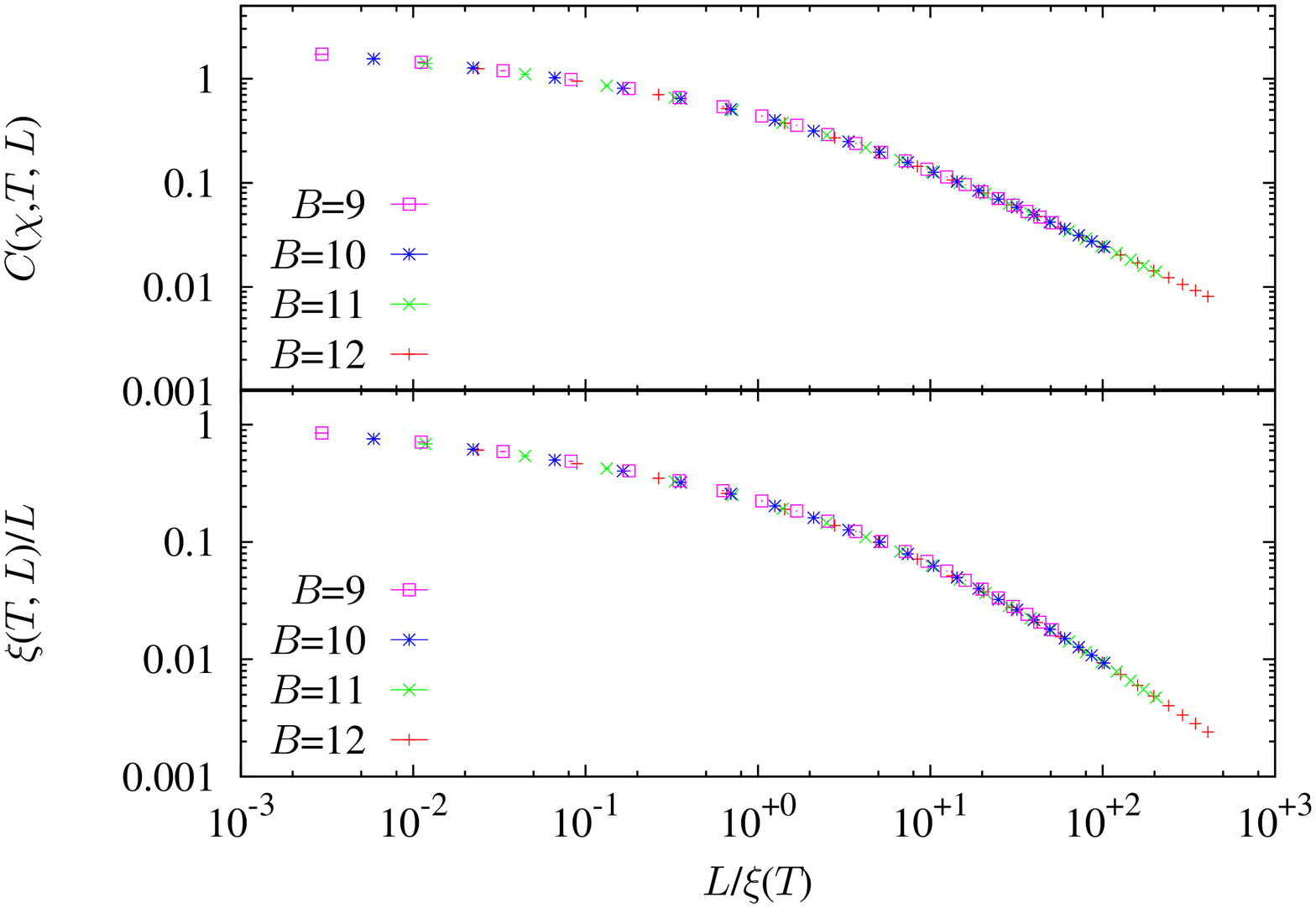}
\caption{(color online) Test on the scaling functions $f_\chi$ and $f_\xi$ for
  $\rho=9/5$. We plot $C(O,T,L) \equiv x^{y_O/\nu} f_O(x)$ versus
  $1/x$ ($O=\xi, \chi$). Notice that $C(\xi,T,L)=\xi(T,L)/L$. }
\label{fig:fig2r1.8b}
\end{figure}

We show in Fig. \ref{fig:fig2r1_8} the scaling behavior of $\xi_2$ and
$\chi$. By fitting the tails, taking into account the statistical
error in both variables, we find $f_{\xi_2}(x)\sim x^{-0.89(5)}$ and
$f_\chi(x) \sim x^{-0.70(3)}$.  These results are to be compared with
$f_{\xi_2}(x)\sim x^{-1}$ and $f_\chi(x) \sim x^{-0.8}$. Once again
the scaling exponents turn out to be underestimated. To gain a deeper
insight on this issue, we, therefore, plot $C(O,T,L)$ versus
$L/\xi(T)$ in Fig. \ref{fig:fig2r1.8b} obtaining finite extrapolated
values as $L/\xi(T) \to 0$.

\begin{figure}[t!]
\centering
\includegraphics[width=.34\textwidth, angle=270]{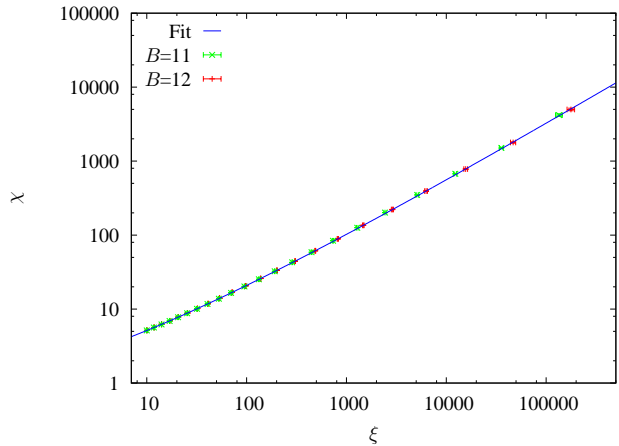}
\caption{(color online) Extrapolated $\chi$ vs. $\xi_2$ for $\rho=9/5$.  Notice the
  two sets of points: one corresponds to the extrapolation of the
  $B=12$-data (red) and the other one to the extrapolation of the
  $B=11$-data (green).}
\label{fig:fig3r1_8}
\end{figure}

Using the $F_O$ and $F_{\xi_2}$ functions (see
Fig. \ref{fig:fig1r1_8}) we can extrapolate the finite volume
correlation length and susceptibility to the thermodynamic limit. In
Fig. \ref{fig:fig3r1_8} we present our results for the infinite volume
susceptibility.  We have fitted the data shown in
Fig. \ref{fig:fig3r1_8} to Eq. (\ref{fit_chi}), and we have obtained
(discarding data with $\xi_2<15$) $\eta=1.221(15)$ and
$\Delta=0.36(7)$ ($\chi^2/\mathrm{d.o.f.}=4.5/14$),\footnote{A pure power
  law fit $\chi= A \xi_2^{2-\eta}$ yields $\eta=1.242(7)$, neglecting
   the $\xi_2$-errors.  Taking into account the statistical uncertainty on
  $\xi_2$, as well as in $\chi$, we find $\eta=1.239(17)$. In both cases the
  quality of the fit is really good. As in $\rho=5/3$, the error on
  $\eta$ has also been doubled.} while, assuming $\eta=3-\rho=1.2$, we obtain $\Delta=0.30(1)$
($\chi^2/\mathrm{d.o.f.}=5.7/15$).

\begin{figure}[t!]
\centering
\includegraphics[width=.34\textwidth, angle=270]{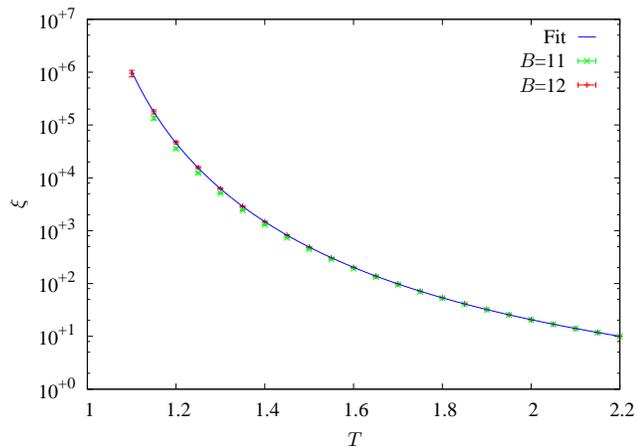}
\caption{(color online) $\rho=9/5$. Extrapolated $\xi_2$ vs. $T$. We have plotted our
  best fit using Eq. (\ref{fit_xi}) (see the text). Notice the two
  sets of points: one corresponds to the extrapolation of the
  $B=12$-data (red) and the other one to the extrapolation
  of the $B=11$-data (blue). The best fits is marked using
  a continuous green line.}
\label{fig:fig4r1_8}
\end{figure}

The final step is the analysis of the correlation length.  By fitting
the data to Eq. (\ref{fit_xi}) (see Fig. \ref{fig:fig4r1_8}) we obtain
$T_c=0.961(8)$, $\nu=5.8(1)$ and $\theta=2.67(6)$
($\chi^2/\mathrm{d.o.f.}=16.9/18$ with $T\le 2.3$). Notice that
$\theta/\nu=0.46(1)$, roughly compatible with the two estimates of
$\Delta$.

As an additional test of the extrapolation procedure, we show in
Figs. \ref{fig:fig3r1_8} and \ref{fig:fig4r1_8} the infinite volume
data from system sizes up to $B=11$ and up to $B=12$: for this value
of $\rho$ data turn out to be statistically compatible.

Finally, we can compare these results with the results obtained using the
behavior of the non-zero Fourier momenta of the spin glass correlation
function:~\cite{Leuzzi11} $T_c=1.060(7)$.

\subsection{Critical behavior at the critical 
threshold exponent $\rho_{\rm L}=2$}

\begin{figure}[t!]
\centering
\includegraphics[width=.34\textwidth, angle=270]{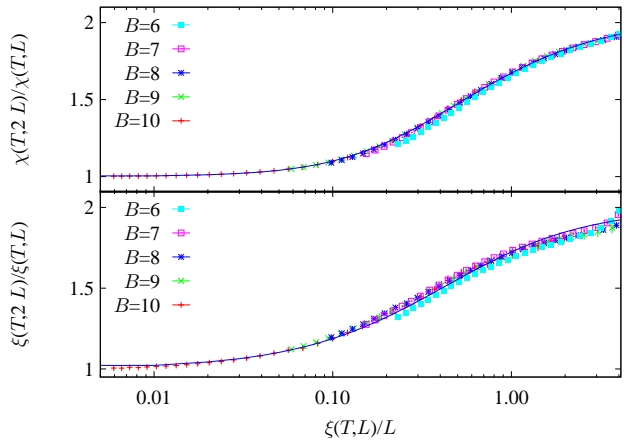}
\caption{(color online) $\rho=2$. FSS functions $F_\xi$ (below) and $F_\chi$
  (above). We plot data for $L=2^{6,7,8,9,10,11}$. We mark with the continuous line
the fit to Eq. ~(\ref{scaling}). }
\label{fig:fig1_r2}
\end{figure}

As a last point we study numerically the model right at the value of $\rho$
corresponding to the lower critical dimension.  In
Fig. \ref{fig:fig1_r2} we again test the Finite Size Scaling Ansatz in the
form of Eq. (\ref{FSS3}). We can also see that, except for the $L=2^6$
system, which suffers stronger scaling corrections, all the data for
larger lattice sizes lie on the same universal curve both for the
susceptibility (top panel) and the correlation length (bottom
panel). The next step has been to parameterize the two universal
functions by means of numerical interpolation. The fits proposed in
references [\onlinecite{Caracciolo95},\onlinecite{Palassini99}] do not
work for $\rho=2$. We have found, though,  that a simple seventh- or
eight-degree cubic spline polynomial fit works well for both
observables. In addition, also fits following Eq. (\ref{scaling}) work quite
well (for $F_\xi$, $\chi^2/\mathrm{d.o.f.}=23.1/34$, and for $F_\chi$,
  $\chi^2/\mathrm{d.o.f.}=16.4/33$, again discarding the $x$-error bars). 
We present, in the following, the outcome of extrapolations according to
Eq.~(\ref{scaling}).

\begin{figure}[t!]
\centering
\includegraphics[width=.34\textwidth, angle=270]{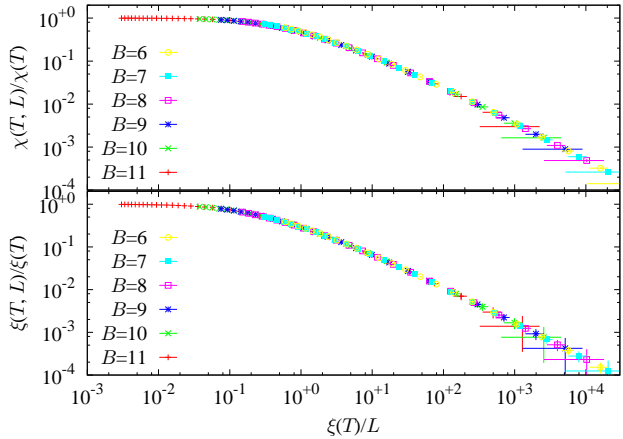}
\caption{(color online) FSS functions $f_\xi$ (below) and $f_\chi$ (above) for $\rho=2$.
We plot data for sizes $L=2^{6,7,8,9,10,11}$.}
\label{fig:fig2_r2}
\end{figure}

\begin{figure}[t!]
\centering
\includegraphics[width=.34\textwidth, angle=270]{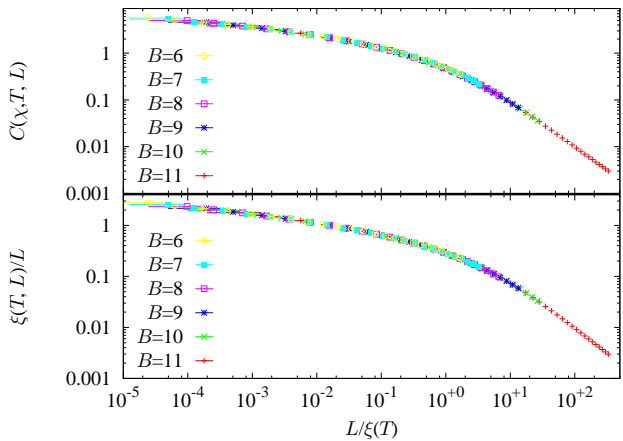}
\caption{(color online) Test on the scaling functions $f_\chi$ and $f_\xi$ for $\rho=2$. We
  plot $C(O,T,L) \equiv x^{y_O/\nu} f_O(x)$ versus $1/x$ ($O=\xi,
  \chi$). Notice that $C(\xi,T,L)=\xi(T,L)/L$.}
\label{fig:fig2_r2b}
\end{figure}

Once again, we check if Eq. (\ref{FSS1}) holds. We present this test in
Fig. \ref{fig:fig2_r2}. We can see that all the points, even those at $L=2^6$,
are lying on the same universal curves (top panel for the susceptibility and
bottom panel for the correlation length).  By fitting the tails we obtain
$f_{\xi_2}(x) \propto x^{-0.86(15)}$ and $f_{\xi_2}(x) \propto x^{-0.87(6)}$
(taking into account the error bars in both axes).  One should expect that
both scaling functions behave as $x^{-1}$, assuming that the relation
$\eta=3-\rho$ is valid down to $\rho=2$. We, thus, repeated the analysis in
term of $C(O,T,L)$, cf. Eq. (\ref{def:COLT}), and the results are plotted in
Fig. \ref{fig:fig2_r2b}: one can see the expected behavior for small values of
$L/\xi(T)$ (i.e. reaching a constant value).

\begin{figure}[t!]
\centering
\includegraphics[width=.34\textwidth, angle=270]{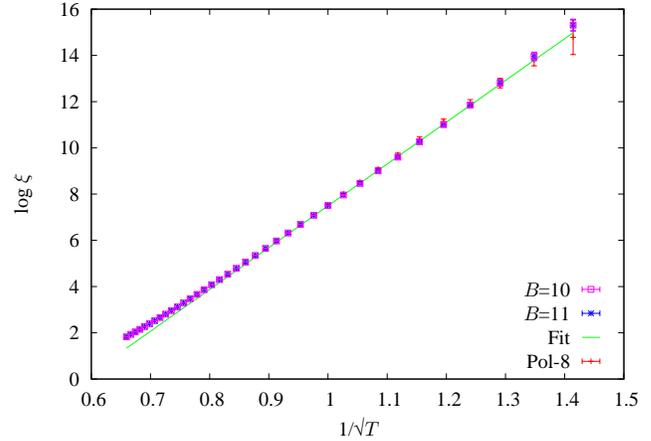}
\caption{(color online) Extrapolated $\log \xi_2$ vs.~$1/\sqrt T$ for $\rho=2$. We
  show points from the extrapolations of sizes up to $B=10$ and up to
  $B=11$, using Eq. (\ref{scaling}). We also show the extrapolated
  points using the alternative fit by means of an eight degree cubic
  spline polynomial (on data up to $B=11$). The three extrapolations
  turn out to be in good agreement.  For small values of the
  temperature the behavior (marked by ``Fit'' in the plot) is linear.}
\label{fig:fig3_r2}
\end{figure}

\begin{figure}[t!]
\centering
\includegraphics[width=.34\textwidth, angle=270]{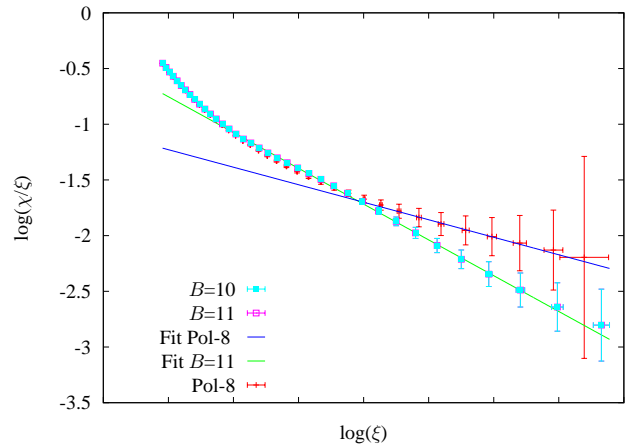}
\caption{(color online) Extrapolated $\log(\chi/\xi)$ vs.~$\log(\xi)$ for $\rho=2$,
  where logarithmic corrections can be appreciated.  The large
  correlation length interpolation over points obtained by means of a
  cubic spline extrapolation (data set of sizes up to $B=11$) is
  consistent with a power law decay with exponent $-0.08(4)$.  A
  even better power-law estimate is obtained using the extrapolation
  Eq. (\ref{scaling}) on the same simulation data and yielding a decay
  exponent of or -0.16(2). Notice the consistency among all three
  extrapolations used.}
\label{fig:fig4_r2}
\end{figure}

The extrapolated correlation length and susceptibility values to the
thermodynamic limit are plotted in Figs. \ref{fig:fig3_r2},
\ref{fig:fig4_r2}. There we show the interpolations performed by means
of Eq. (\ref{scaling})
 for data sizes up to $B=10$ and up to $B=11$ and also by means of the cubic
spline fit.  Our data for $\xi_2$ are well fitted by a law like 
\begin{equation}
  \label{xi_T}
\xi(T,\infty) \propto \exp\left(\frac{a}{\sqrt{ T}}\right) \,,
\end{equation}
where $a=18.1(2)$ ($\chi^2/\mathrm{d.o.f.}=4.45/9$).
The simulated numerical data are not
compatible, though, with the law $\xi \propto \exp(-a\log T/T^2)$
suggested by Moore,~\cite{Moore10} (at least for $T\ge 0.5$),
but it is worth reminding that the fully connected version studied
by Moore and the diluted version we simulate may have a different critical
behavior at $\rho=2$.~\cite{Weigel} 

Finally, we analyze the relationship between susceptibility and
correlation length. From a naive theoretical point of view, from the
law $\chi\propto \xi^{2-\eta}$, we should expect a relation as
$\chi\propto \xi$ in $\rho=2$, assuming $\eta=1$. This linear
relation is possibly modified by logarithmic corrections. In
Fig. \ref{fig:fig4_r2} we plot $\log(\chi/\xi)$ versus $\log(\xi)$.
One can see that finite size corrections to the leading behavior are
there, though it is rather difficult to precisely determine their
nature. Data are, indeed, consistent with logarithmic corrections, as
well as power-law corrections with small exponents.  The latter are
estimated using  data set of sizes up to $B=11$, 
either with an exponent  $-0.08(4)$, using a large $\xi_2$ interpolation
over points obtained by means of a cubic spline extrapolation, 
or with an exponent $-0.16(2)$, by means of  Eq. (\ref{scaling}).
 With  the latter kind of behavior, one has $\chi\propto
\xi^{1-0.16(2)}$, a bit different from the naive theoretical
prediction. In any case,  such small correction $\xi^{-0.16(2)}$
is very hard to be distinguished from a logarithmic correction.

\section{Discussion}

\begin{figure}[t!]
\centering
\includegraphics[width=.34\textwidth, angle=270]{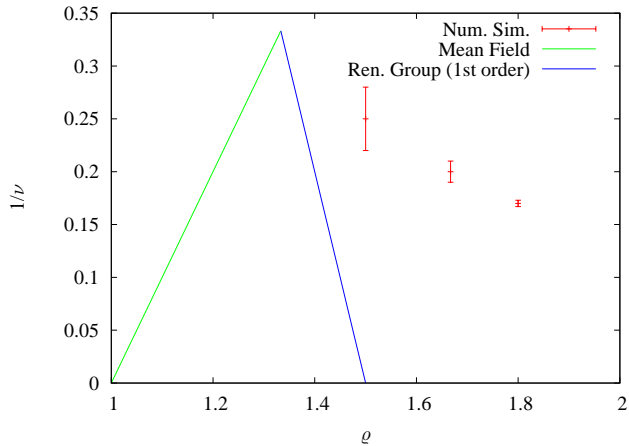}
\caption{(color online) Behavior of $1/\nu$ as a function of $\rho$. The green
  straight line is the MF prediction ($1/\nu=\rho-1)$, the blue line
  is the results of the first order $\epsilon$-expansion and the
  points are from numerical simulations: the two rightmost points are
  from this work.}
\label{fig:nu}
\end{figure}

In Fig. \ref{fig:nu} we have plotted the behavior of $1/\nu$ as a function of
$\rho$. Together with our numerical estimates, we have drawn the mean field
prediction ($1/\nu=\rho-1$), which is valid for $\rho<4/3$ and the prediction
from a first order renormalization group (RG) calculation, that should be
valid very close to $\rho=4/3$. Since for $\rho=2$ we expect $1/\nu=0$, the
decrease should be very fast and likely incompatible with the linear behavior
$1/\nu \propto (2-\rho)$, predicted in Ref.~\onlinecite{Moore10}.  Such a
difference may be due to a possibly different critical behavior between the
fully-connected and the diluted versions of the model.~\cite{Weigel} However
another possibility is that one of the approximations made in
Ref.~\onlinecite{Moore10} in order to solve the RG equations is too crude:
actually the author of Ref.~\onlinecite{Moore10} warns the reader, just after
Eq.~(35), that the approximation made is not valid close to $T_c$ for $\rho<2$
(which is exactly the region we are studying).

The behavior of the correlation length that  we have found is consistent 
with the following  renormalization flow of the temperature
\begin{equation}
\frac{d T}{d l} \propto T^{3/2}\,\,\, \mathrm{as} \,\,\, T\to 0\,,
\end{equation}
whereas the phenomenological renormalization of Ref.~\onlinecite{Moore10}
predicts a different leading behavior like
\begin{equation}
\frac{d T}{d l} \propto \frac{T^3}{\log T} \,\,\, \mathrm{as} \,\,\, T\to 0\,,
\end{equation}
not compatible with our numerical data.
This is another motivation to reconsider the approximation made in
Ref.~\onlinecite{Moore10}.
 
\section{Conclusions}

We have numerically revisited the one dimensional bond diluted Levy Ising spin
glass.\cite{Leuzzi08,Leuzzi09} In particular we have focused in the less
explored region of power-law decaying interaction with large power-law
exponents, not compatible with a mean-field critical behavior.  Being
$\rho=4/3$ the mean-field threshold, we have been analyzing data for the
critical behavior of systems with $\rho=5/3, 9/5$ and $2$. The latter being
the exponent of the long-range model whose critical behavior is at zero
temperature.  Through a careful finite size scaling analysis we have been able
to extrapolate, to infinite volume, refined susceptibility and correlation
length scaling behaviors.  These results allows us to test analytical
predictions for the behavior at the lower critical dimension, corresponding to
$\rho=2$, as the renormalization flow towards the zero temperature fixed point
and the correlation length behavior in temperature. For the critical
temperature flow our data are not compatible with the picture obtained in
Ref. [\onlinecite{Moore10}] (see Ref. [\onlinecite{Palassini99}] for a similar
discussion in the finite dimensional model). For the $\xi(T)$ behavior our
data are compatible with Eq. (\ref{xi_T}) and not with the law proposed in
Ref.  [\onlinecite{Moore10}]. Quite generally, the methods used in this paper
are very suitable for studying models near their lower critical dimension.

\section{Acknowledgments}

This work was partially supported by the Ministerio de Ciencia y
Tecnolog\'{\i}a (Spain) through Grant No. FIS2013-42840-P, by the Junta de
Extremadura (Spain) through Grant No. GRU10158 (partially founded by
FEDER), by European Union through Grant No. PIRSES-GA-2011-295302,
by European Research Council (ERC) through grant agreement No. 247328, by
the Italian Ministry of Education, University and Research under
the Basic Research Investigation Fund (FIRB/2008) through grants
No. RBFR08M3P4 and RBFR086NN1, and under the
PRIN2010 program, grant No. 2010HXAW77-008 and by the People
Programme (Marie Curie Actions) of the European Union's Seventh
Framework Programme FP7/2007-2013/ under REA grant agreement
n. 290038, NETADIS project.

\end{document}